\documentclass[doublecol,figures]{epl2}

\title{Chain Formation by Spin Pentamers in $\eta$-Na$_9$V$_{14}$O$_{35}$}
\shorttitle{Chain Formation by Spin Pentamers in
$\eta$-Na$_9$V$_{14}$O$_{35}$}

\author{D.~V.~Zakharov\inst{1} \and H.-A.~Krug~von~Nidda\inst{1} \and J.~Deisenhofer\inst{1} \and
F.~Schrettle\inst{1} \and G.~Obermeier\inst{2} \and S.~Horn\inst{2}
\and A.~Loidl\inst{1}}
\shortauthor{D.~V.~Zakharov \etal}

\institute{
  \inst{1} Experimental Physics V, Center for Electronic Correlations and Magnetism,
  University of Augsburg -- Universit\"{a}tsstra{\ss}e 2, 86135 Augsburg, Germany\\
  \inst{2} Experimentalphysik II, Institut f\"ur Physik, Universit\"{a}t
  Augsburg -- Universit\"{a}tsstra{\ss}e 2, 86135 Augsburg, Germany
}


\abstract{ The nature of the gapped ground state in the
quasi-one-dimensional compound $\eta$-Na$_9$V$_{14}$O$_{35}$ cannot
easily be understood, if one takes into account the odd number of
spins on each structural element. Combining the results of specific
heat, susceptibility and electron spin resonance measurements we
show that $\eta$-Na$_9$V$_{14}$O$_{35}$ exhibits a novel ground
state where multi-spin objects build up a linear chain. These
objects - pentamers - consist of five antiferromagnetically arranged
spins with effective spin 1/2. Their spatial extent results in an
exchange constant along the chain direction comparable to the one in
the high-temperature state. }

\begin{document}

\maketitle

%%%%%%%%%%%%%%%%%%%%%%%%%%%%%%%%%%%%%%%%%%%%%%%%%%%%%%%%%%%%%%%%%%%%%%%%%%%%%%%%%%%%%%%%%%%%%%%%%%%%%%%%%%%

\section{Introduction}

The complex magnetic properties of transition-metal oxides are a
consequence of the strong interplay of charge, spin, lattice, and
orbital degrees of freedom \cite{Imada98}. The physics becomes
particularly fascinating in systems with lower dimensionality, where
quantum fluctuations suppress the conventional magnetic order
\cite{Mermin66} in favour of exotic ground states like the
spin-Peierls dimerization in $S = 1/2$ chains \cite{Pytte74} or the
Haldane-gap formation in integer-spin chains \cite{Haldane83}.

Among all low-dimensional transition-metal oxides one can single out
the vanadium bronzes which acquired a paradigmatic status because of
the variety of phenomena displayed by these systems. The possibility
to tune the vanadium valence between V$^{4+}$ ($3d^1$) and V$^{5+}$
($3d^0$) allows the realization of a multitude of spin-1/2 systems
with strong quantum effects. Moreover, the rich structural chemistry
of these systems, where the V ions can occur in pyramidal,
tetrahedral, or octahedral coordination, gives rise to the formation
of very interesting chain- and ladder-like structures
\cite{Evans90,Kanke90,Ivanshin03}. The quarter-filled spin ladder
$\alpha'$-NaV$_{2}$O$_{5}$ \cite{Isobe96,Eremin06} and the
one-dimensional metal $\beta$-Na$_{1/3}$V$_{2}$O$_{5}$
\cite{Yamada99,Heinrich04} are the most intensively studied members
of this series. The former system reveals charge ordering with the
opening of a spin gap similar to a spin-Peierls transition, the
latter shows a metal-to-insulator transition and even becomes
superconducting under pressure.

\begin{figure}
\centering
\includegraphics[width=80mm]{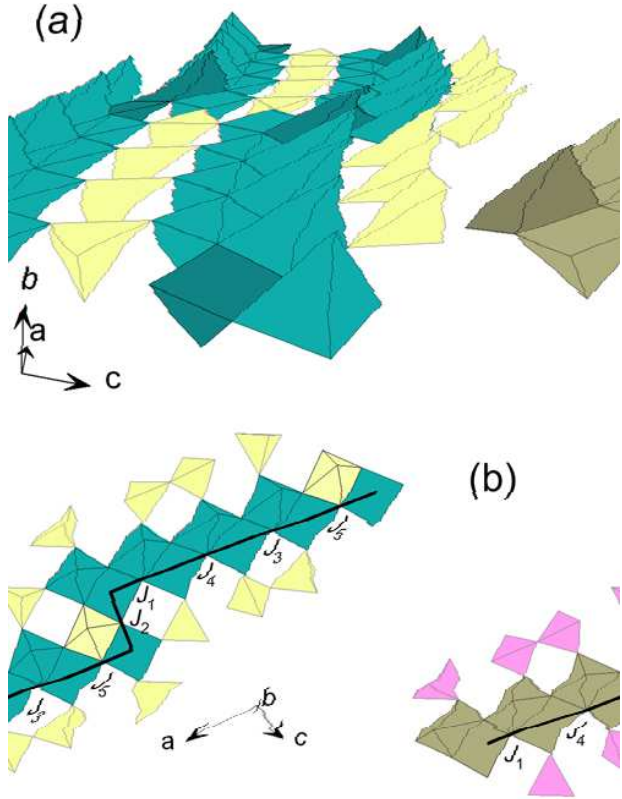}
\caption{(color online) (a): Part of the high-temperature ($T >
100$~K) crystal structure of $\eta$-Na$_9$V$_{14}$O$_{35}$. (b):
Projection of the low temperature crystal structure onto the
$(ac)$-plane. The cyan (dark cyan) pyramids correspond to V$^{4+}$
(V$^{4.5+}$) ions surrounded by the O$^{2-}$ ions, the light yellow
VO$_4$ tetrahedra contain non-magnetic V$^{5+}$ ions. $J_i$ denote
exchange constants between adjacent V ions along the chain.
\{$a',b',c'$\} denotes the local coordinate system of the VO$_5$
pyramids. }\label{Structure}
\end{figure}

In this Letter we focus on a vanadium oxide bronze with higher Na
concentration, $\eta$-Na$_{9}$V$_{14}$O$_{35}$.
%with the stoichiometric formula Na$_9$V$_{14}$O$_{35}$.
This system came recently into the focus as a unique example of a
low-dimensional spin-gap system \cite{Isobe99}. The temperature
dependence of the susceptibility indicates the opening of a spin gap
below 30~K, while at high temperatures following the Bonner-Fisher
(BF) law \cite{Bonner64} with $J \simeq 190$~K. Its low-temperature
behavior, however, cannot be described by any known model for
spin-gap systems \cite{Isobe99}. The room-temperature structure of
$\eta$-Na$_9$V$_{14}$O$_{35}$ is monoclinic with space group $P2/c$.
As depicted in Fig.~\ref{Structure} the structure consists of double
chains of the corner-sharing VO$_4$ pyramids running along the $a$
direction, exhibiting, however, a crystallographic step at every
five VO$_5$ unit. The edge-sharing exchange interactions inside
these chains can be neglected \cite{Koo07}. The double chains are
bridged by VO$_4$ tetrahedra, containing the non-magnetic V$^{5+}$
ions, in the $(ac)$-plane to form the V$_2$O$_5$ layers which are
mediated by the Na ions along the $b$ axis. These chains excluding
the steps are structurally similar to those of
$\alpha'$-NaV$_2$O$_5$, but have a larger average vanadium valence.
Eight out of ten V ions are tetravalent except for two neighboring V
ions on the structural step, which share one electron and,
therefore, have an average valence $+4.5$ at high temperatures.

Based on a spin-dimer analysis it was suggested that the system can
be understood in terms of 10-node rings of V$_{10}$O$_{30}$, with
each ring opening a spin gap \cite{Whangbo00}. This scenario has
been discarded by Duc \etal \cite{Duc04} arguing that a spin gap can
only occur, if the number of magnetic sites per unit cell is even.
With its nine magnetic sites per unit cell this is evidently not the
case for $\eta$-Na$_9$V$_{14}$O$_{35}$. The authors attempted to
explain the occurrence of a spin gap with the observed charge
ordering superstructure on the two V$^{4.5+}$ sites at the
structural step below 100~K, which subsequently leads to a larger
unit cell doubled along the $b$-axis and 18 spins per unit cell.
However, the weak exchange coupling between adjacent layers along
the $b$ axis \cite{Whangbo00} casts doubts upon the effectiveness of
such a mechanism for a complete spin-gap formation. Therefore, the
inter-chain exchange is thought to be the driving force behind the
spin gap opening \cite{Duc04}. Based on a quantitative analysis of
all possible exchange paths Koo and Whangbo \cite{Koo07} recently
proposed that the magnetic structure of
$\eta$-Na$_9$V$_{14}$O$_{35}$ is build up by chains of 12-spin-rings
alternating with chains of 10-spin rings implying the existence of
two different spin gaps. Alone, no experimental evidence for such a
scenario has been reported and the challenging task remains to
consistently describe the spin-gap-like nature of the magnetic
system with structural elements containing an odd number of spins.

\begin{figure}
\centering
\includegraphics[width=\linewidth]{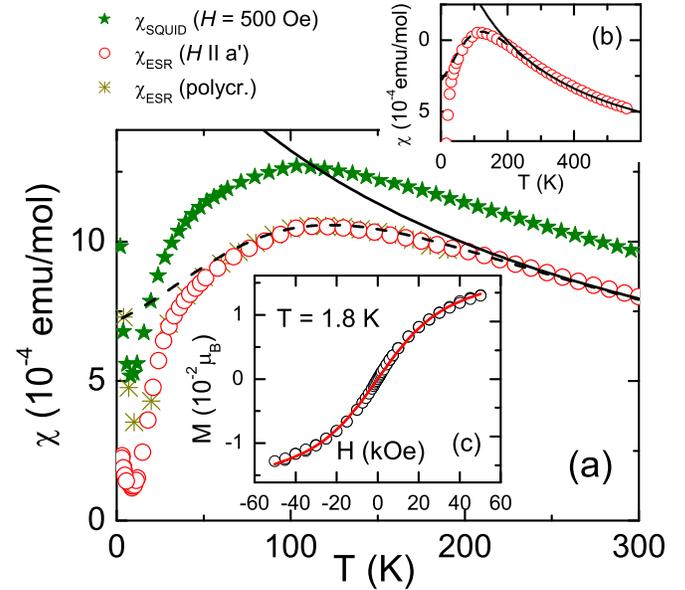}
\caption{(color online) Main frame (a): Temperature dependence of
the spin susceptibility $\chi$ measured both by SQUID magnetometry
and ESR in Na$_9$V$_{14}$O$_{35}$. The black dash line represents
the prediction of the BF model \cite{Bonner64} with $J/k_{\rm B} =
190$~K and nine spins on each structural element, the solid line --
of the Curie-Weiss model with $\Theta = - 200$~K. The enlarged view
of the low-temperature data is presented in Fig.~\ref{PRL_LowTsysc}.
Inset (b): Temperature dependence of $\chi_{ESR}$ with $H
\parallel a'$ at high temperatures. Inset (c): Magnetization $M(H)$ at $T = 1.8$~K.
The line represents a fit by a sum of a Brillouin function and a
temperature independent term $M_0 = \chi_0 H$.} \label{PRL_AllSysc}
\end{figure}

In this work we resolve this ostensible contradiction based on the
analysis of the magnetic and thermal properties of
$\eta$-Na$_{9}$V$_{14}$O$_{35}$. We interpret the magnetic structure
to be dominated by continuous spin chains along the $a$ axis and as
a result the opening of the spin gap at low temperatures remains
\emph{incomplete}. Our alternative model for the magnetic ground
state, namely a spin chain consisting of spin-pentamer building
blocks, allows for a consistent description of the properties of
$\eta$-Na$_{9}$V$_{14}$O$_{35}$.

\section{Sample preparation and experimental details}

The single crystals of Na$_9$V$_{14}$O$_{35}$ were grown in a two
step process: Firstly, pellets of a nearly stoichiometric mixture of
high purity NaVO$_3$ and VO$_2$ were pressed and heated in an
evacuated quartz tube at 650$^{\circ}$C for four days. Then the
material was heated above the melting temperature and, in a
temperature gradient, was cooled down at a cooling rate of
7$^{\circ}$C/h. Debye-Scherrer x-ray diffraction and Laue
diffraction showed the material to be single phase. The
heat-capacity and susceptibility measurements have been performed
with a commercial physical properties measurement system (PPMS) and
a SQUID magnetometer (MPMS5), both from Quantum Design, for
temperatures 1.8~K~$<T<$~300~K. The Electron Spin Resonance (ESR)
experiments have been carried out with a Bruker ELEXSYS E500
CW-spectrometer at X-band frequency ($\nu \approx$ 9.4~GHz). The
details of the experimental ESR set-up can be found elsewhere
\cite{Zakharov05}. The obtained ESR spectra are in good agreement
with previously published results \cite{Duc04}. For the
heat-capacity measurements polycrystalline samples with appropriate
mass were used.

\section{Magnetic Susceptibility}

We will start with the analysis of the spin susceptibility of the
system as measured by ESR. To calculate the absolute values of the
spin susceptibility $\chi_{\rm ESR}$ in
$\eta$-Na$_{9}$V$_{14}$O$_{35}$ from the double-integrated ESR
spectra, we compared the measured intensities to the ESR signal of
the reference compound Gd$_2$BaCuO$_5$. The $dc$-susceptibility of
this system follows a Curie-Weiss law $\chi_{\rm ESR} = C / (T -
\Theta)$ at $T > 100$~K, the Curie constant of which corresponds to
the contribution of all Gd ions \cite{Goya96}. For this comparison
we used the high-temperature ESR data (for $T
> 300$~K) of Na$_9$V$_{14}$O$_{35}$ which can be described by a Curie-Weiss
law with $\Theta = -200$~K (see Fig.~\ref{PRL_AllSysc}(b)).

The obtained effective magnetic moment $\mu_{\rm eff} = 5 \,
\mu_{\rm B}$ coincides well with the magnetic moment $\mu_{\rm
theor} = 5.1 \, \mu_{\rm B}$ expected for nine magnetic
V$^{4+}(3d^1)$ ions per unit cell. Therefore, we conclude that all
vanadium ions participate in the generation of the ESR signal in
Na$_9$V$_{14}$O$_{35}$.

\begin{figure}
\centering
\includegraphics[width=\linewidth]{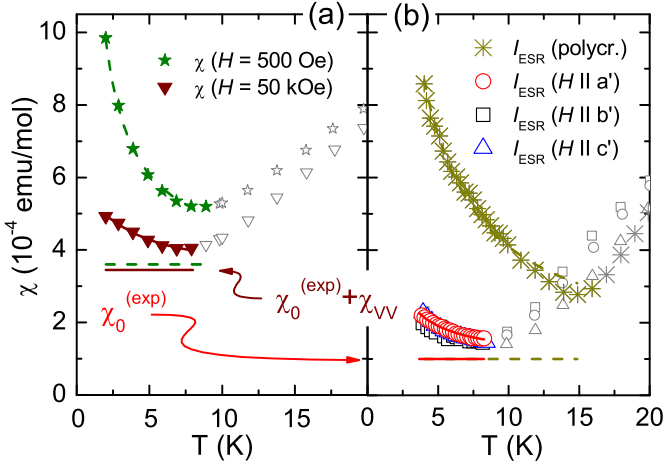}
\caption{(color online) Magnetic susceptibility of
Na$_9$V$_{14}$O$_{35}$ at low temperatures. The lines represent fits
by a sum of a Brillouin function and a temperature independent term
$\chi_0$ (shown separately as a constant contribution). (a):
Susceptibility in a polycrystal measured by SQUID magnetometry in a
magnetic field $H=500$~Oe (green stars) and $H=50$~kOe (dark-red
triangles). (b): Intensity of the ESR signal $I_{\rm ESR}$ in a
polycrystal (dark-yellow cross stars) and in a single crystal for
the magnetic field applied along the three crystallographic axes.}
\label{PRL_LowTsysc}
\end{figure}

\begin{figure}
\centering
\includegraphics[width=\linewidth]{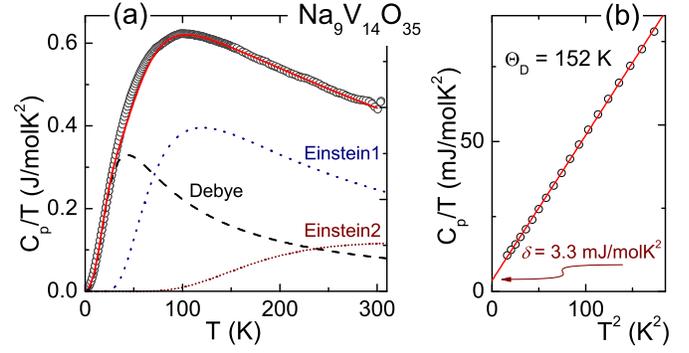}
\caption{(color online) (a): Heat capacity of
$\eta$-Na$_{9}$V$_{14}$O$_{35}$ plotted as $C_p/T$ vs $T$. The
contribution of acoustical Debye-phonons is indicated by the dashed
black line, contributions of two Einstein modes with $\Theta_{\rm
E1} = 310$~K and $\Theta_{\rm E2} = 850$~K by the blue dotted and
the wine short-dotted line, respectively. The sum of all
contributions (red solid line) represents the best fit to the
experimental results. Frame (b) emphasizes the Debye and the
spin-chain contribution at low temperatures.} \label{PRL_C}
\end{figure}

Fig.~\ref{PRL_AllSysc} shows the temperature dependence of
$\chi_{\rm ESR}$ measured both in a single crystal (with $H
\parallel a'$) and in a polycrystalline sample. Below room
temperature $\chi_{\rm ESR}$ does not follow the Curie-Weiss law
anymore. Such kind of behavior is typical for one-dimensional spin
systems and can be fitted by using the Bonner-Fisher (BF) model
\cite{Bonner64}. Using the integral of the isotropic exchange
between neighboring spins $J/k_{\rm B} \approx 190$~K and assuming
nine spins per unit cell, we were able to describe the magnitude and
the temperature dependence of the susceptibility down to $T \sim
50$~K.

Let us now compare the spin susceptibility obtained from the
intensity of the ESR absorption signal $\chi_{\rm ESR}$ with the
magnetic $dc$-susceptibility $\chi_{\rm SQUID}$ measured by SQUID
magnetometry. As one can see in Fig.~\ref{PRL_AllSysc}, both of them
have the same temperature dependence with a broad maximum at around
100~K, in accordance with Refs.~\cite{Isobe99,Duc04}. The
$dc$-susceptibility is larger because of the Van Vleck (VV) orbital
contributions to the static susceptibility, which do not affect the
spin susceptibility measured by ESR \cite{Yosida96}.

\begin{figure*}
\begin{center}
\includegraphics*[width=0.8\linewidth]{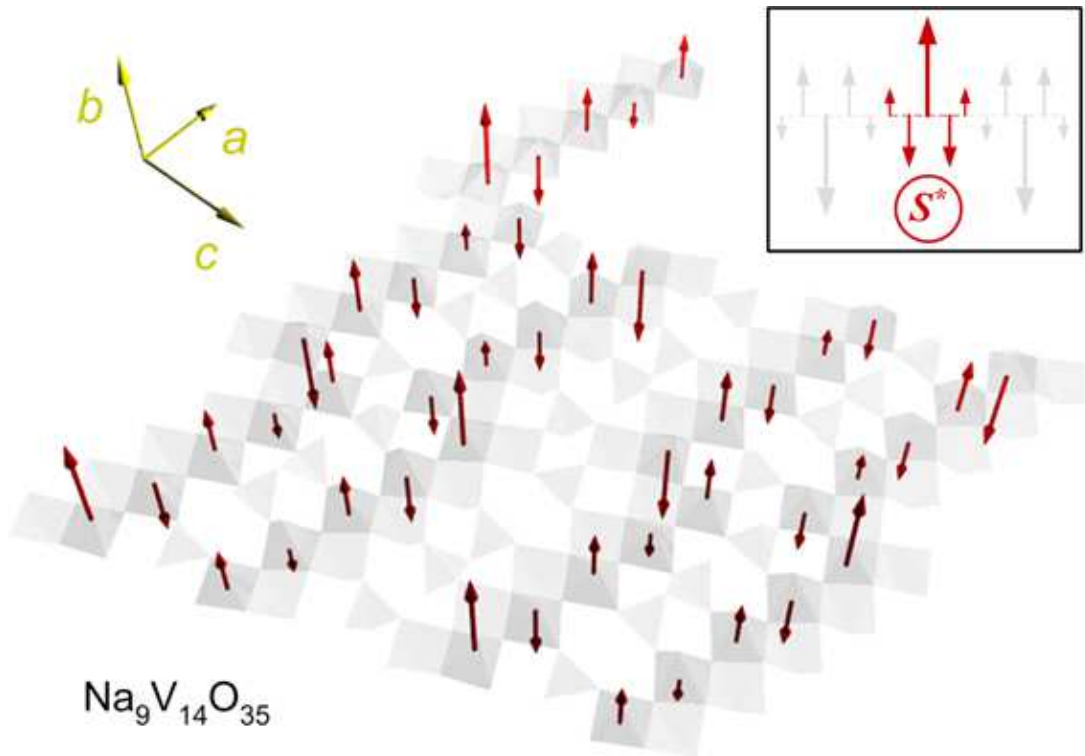}
\caption{(color online) Schematic spin structure of the proposed
ground state in $\eta$-Na$_{9}$V$_{14}$O$_{35}$ at $T < 10$~K. The
inset shows the spin structure which appears near an unpaired spin.
The arrows represent the average spin projections at the lattice
sites. } \label{PRL_Multi}
\end{center}
\end{figure*}

Below 50~K the data decrease faster than the BF law suggesting a
dimerization of the spins, but finally turn up again below 15~K.
Such an increase of the susceptibility to lower temperatures usually
arises owing to a small amount of paramagnetic impurities present in
the samples. The measurements of the magnetization $M(H)$
(Fig.~\ref{PRL_AllSysc}(c)) and of the susceptibility in a higher
field $H = 50$~kOe (Fig.~\ref{PRL_LowTsysc}(a)) show that this
paramagnetic contribution starts to saturate already above $H
\approx 10$~kOe. The fit curves given in the pictures represent the
best fit of these data achieved by a sum of a Brillouin function
\begin{equation} \label{BrF}
M \propto (2S+1) \textrm{cth}\left(\frac{2S+1}{2S} \frac{\gamma S
H}{k_{\rm B}T}\right) - \textrm{cth}\left(\frac{1}{2S}\frac{\gamma S
H}{k_{\rm B}T}\right) \nonumber
\end{equation}
($\gamma$ denote the gyromagnetic ratio, $k_{\rm B}$ is the
Boltzmann constant, the spin $S = 1/2$) and a temperature
independent contribution $M_0 = \chi_0 H$. An alternative
description of the low-temperature susceptibility with a Curie-Weiss
law taking into account a residual magnetic interaction felt by
impurities but without any constant contribution was proposed in
Ref.~\cite{Duc04}. This allowed to fit $\chi(T)$ at low fields only,
but we did not achieve a reasonable description of the field
dependence of the magnetization and of the high-field susceptibility
in terms of this approach.

The Brillouin contribution is a sign of free impurity spins, while
the temperature independent susceptibility $\chi_0$ may, in
principle, originate from four possible contributions, namely, (i)
saturated ferromagnetic moments, (ii) Pauli paramagnetism, (iii) the
Van Vleck orbital
paramagnetism, and (iv) the spin chain contribution:\\
The paramagnetic behavior of the magnetization $M(H)$
(Fig.~\ref{PRL_AllSysc}(c)) allows to exclude the presence of
ferromagnetic moments in the sample. To check for Pauli
paramagnetism we performed dielectric measurements of
$\eta$-Na$_9$V$_{14}$O$_{35}$ that revealed values of the electric
conductivity less than 10$^{-12}~\Omega^{-1}$cm$^{-1}$ for $T <
50$~K and, hence, allow to neglect a possible electronic
contribution to $\chi_0$. The magnitude of the Van Vleck term
$\chi_{\rm VV}$ can be estimated by comparison with ESR data,
because the intensity of the ESR absorption signal $\chi_{\rm{ESR}}$
is not sensitive to VV contributions. As one can see in
Fig.~\ref{PRL_LowTsysc}(b) the temperature independent background is
indeed considerably smaller for $\chi_{\rm ESR}(T)$, but still
remarkable and amounts to $\chi^{\rm (exp)}_0 \approx 0.95 \cdot
10^{-4}$~emu/mol \cite{Note1}. Consequently, we find ourselves left
with the contribution of a spin-chain system which seems to survive
despite the considerable reduction of the susceptibility at $T \sim
30$~K.

To resolve this challenging task we start out with the analysis of
the spin chain contribution to the magnetic and thermodynamic
properties in the low-temperature
phase:\\
According to Bonner and Fisher \cite{Bonner64}, the magnitude of the
susceptibility at zero temperature $\chi^{\rm (BF)}_0$ is expected
to be about 0.69 of its maximal value $\chi_{\rm max}^{\rm (BF)} =
0.147 \cdot g^2 \mu_{\rm B}^2 N / |J|$. Using $\chi_{\rm max}
\approx 10.6 \cdot 10^{-4}$~emu/mol (Fig.~\ref{PRL_AllSysc}), one
gets $\chi^{\rm (BF)}_0 \approx 7.3 \cdot 10^{-4}$~emu/mol. The
experimental contribution $\chi^{\rm (exp)}_0$, however, amounts to
13\% of this value only.

Let us turn now to the results of the specific heat measurements
which confirm the presence of the spin chain contribution at low
temperatures in $\eta$-Na$_9$V$_{14}$O$_{35}$. Moreover, the
magnitude of this contribution turns out to be in good agreement
with the susceptibility data.

\section{Specific Heat}

Following Bonner and Fisher \cite{Bonner64} a linear increase
$C_{\rm p}(T) = \delta T$ of the specific heat is characteristic for
a linear antiferromagnetic spin-1/2 chain at low temperatures $T \ll
|J|/k_{\rm B}$. The plot of $C_{\rm p}/T$ vs. $T^2$ allows to reveal
this type of behavior and estimate the magnitude of $\delta$.
Fig.~\ref{PRL_C}(b) shows that the specific heat of
Na$_9$V$_{14}$O$_{35}$ at low temperatures $T < 5$~K cannot be
explained by the Debye-contribution $C_{\rm p}(T \ll \Theta_{\rm D})
\propto T^3$ only. Because of the underlying linear contribution the
fit line crosses the ordinate axis not in the origin but at $C_{\rm
p}/T = \delta_{\rm exp} = 3.3$~mJ/molK$^2$. This value is again
smaller than the expected one and amounts to approximately 11.3\% of
the value predicted by the calculations of Bonner and Fisher
$\delta_{\rm BF} = 0.7 k_{\rm B} R / |J| \approx 29.1$~mJ/molK$^2$.

To summarize the experimental findings so far, both the magnetic and
the thermodynamic properties of a linear spin-1/2 chain are present
in Na$_9$V$_{14}$O$_{35}$ despite the partial dimerization of the
spin structure at $T < 50$~K. Only the corresponding parameters
$\chi_0$ and $\delta$ are modified pointing at a nontrivial
distribution of the spin density and the exchange coupling constants
along the chain. In the following we will try to reproduce the
low-temperature spin structure based on the previous results
\cite{Duc04,Koo07}, our experimental data, and the theoretical
calculations \cite{Bonner64}.

\section{Analysis}

The origin of the spin-chain contributions becomes clear, if one
takes into account the results of the spin-dimer analysis by Koo and
Whangboo \cite{Koo07}. According to them, the high-temperature $T >
100$~K magnetic structure can be considered as a
quasi-one-dimensional system consisting of double chains running
along the $a$ direction (see Fig.~\ref{Structure}(a)). At low
temperatures one of the adjacent chains breaks off, whereas the
second one $(J_5-J_3-J_4-J_1-J_2-)_{\infty}$, shown by the black
line in Fig.~\ref{Structure}(b), survives with almost unchanged
exchange parameters $J_i$ along the chain. This chain can be
regarded as a result of linking pentamers with the $J_3-J_4-J_1-J_2$
paths through the paths $J_5$ \cite{Koo07}. The spin pentamer does
not have a spin gap preserving the spin-chain properties down to
lower temperatures. Two of these properties, a temperature
independent contribution to the magnetic susceptibility and a linear
contribution to the specific heat, are clearly seen in the
experiment.

The magnitude of these contributions in the low-temperature phase
can be estimated using the following expressions \cite{Bonner64}:
\begin{equation}\label{EtaRatio}
\chi_0 = \frac{g^2 \mu_{\rm B}^2}{\pi^2} \cdot
\left(\frac{N}{J}\right), \quad \delta = 0.7 \, k_{\rm B}^2 \cdot
\left(\frac{N}{J}\right),
\end{equation}
where $N$ and $J$ denote the number of involved spins and the
effective exchange integral, respectively. The values of these
contributions, $\chi^{\rm (BF)}_0$ and $\delta_{\rm BF}$, estimated
previously, were obtained using the high-temperature values of $N$
(nine spins per structural element) and $J$ (190~K). Whereas the
values of exchange integrals along the chain remain almost
unaffected by the charge-ordering transition at $T \approx 100$~K
\cite{Koo07}, the number of involved spins $N$ might be changed at
lower temperatures and be responsible for the reduced magnitude of
these contributions $\chi^{\rm (exp)}_0$ and $\delta_{\rm exp}$.

All models of the low temperature magnetic ground state proposed so
far \cite{Duc04,Koo07} cannot explain the presence and the magnitude
of $\chi^{\rm (exp)}_0$ and $\delta_{\rm exp}$. However, it can be
understood on the basis of an effect predicted theoretically
\cite{Fukuyama96} in the diamagnetically diluted spin-Peierls system
CuGeO$_3$ and confirmed experimentally in Cu$_{1-x}$Mg$_x$GeO$_3$
\cite{Glazkov05} and Pb(Ni$_{1-x}$Mg$_x$)$_2$V$_2$O$_8$
\cite{Smirnov02}. In these works it was shown that an impurity spin
causes the occurrence of a local magnetization on the neighboring
sites in order to gain exchange energy. The induced magnetic moment
is staggered because of the antiferromagnetic coupling between the
spins and decreases with increasing distance from the impurity spin
as shown schematically in the inset of Fig.~\ref{PRL_Multi}.
However, the net magnetic moment of of the whole structure remains
equal to $1\mu_{\rm B}$. The characteristic length $L$ of this
multi-spin object (the extension of short range antiferromagnetic
ordering) along the chain direction depends on temperature $J S^2
e^{-2 L / \xi} \sim k_{\rm B} T$ and is of the order of the
correlation length $\xi$, which was estimated to be of about ten
interspin distances for the above compounds.

In $\eta$-Na$_{9}$V$_{14}$O$_{35}$ each structural pentamer (five V
ions connected via $J_3-J_4-J_1-J_2$, see Fig.~\ref{Structure})
possesses an odd number of spins. Hence, one of them (every fifth
one) is uncompensated and serves as an impurity spin $S^*=1/2$. By
analogy with the cases described in
Refs.~\cite{Fukuyama96,Glazkov05,Smirnov02}, this magnetic moment
will spread out on the neighboring four spins building spin
pentamers, objects with five antiferromagnetically coupled spins
with effective spin 1/2 (see the inset of Fig.~\ref{PRL_Multi}).
Note that these pentamers overlap strongly because the
characteristic magnetic correlation length is of the order of the
segment length. Therefore, the spin pentamers has to be coherent
along the whole chain (Fig.~\ref{PRL_Multi}). The magnetic and
thermal properties of this effectively antiferromagnetic chain
consisting of such spin pentamers are expected to be quantitatively
different from the original chain because of the considerably
smaller average magnetic moment on each site. But the exchange
energy of this structure with an almost unchanged local exchange
constant can still produce a linear contribution to the specific
heat and a temperature independent contribution to the
susceptibility. The experimental results evidence the presence of
these contributions.

The values of these contributions are supposed to be nine times
smaller as compared to the values expected for two non-dimerized
$S=1/2$ chains (with 9 spins per 10 sites) realized at
high-temperatures: one of the chains is completely broken off at low
temperatures and the effective spin of the second chain becomes five
times smaller. The experimental reduction factors 0.13 for the
susceptibility and 0.113 for the specific heat are in good
accordance with the expected ratio $1/9 \approx 0.111$.

Thus, the spin-gap opening in the low-temperature state of
$\eta$-Na$_{9}$V$_{14}$O$_{35}$ turns out to be incomplete. This
conclusion is supported by a recent polarized Raman-scattering
investigation \cite{Popovic07}. The observed Raman modes become
narrower, change their energy and width, but no new modes were
detected on lowering the temperature down to 10~K, although it is
natural to expect the appearance of new modes in the spin or phonon
channel in case of a conventional charge ordering transition with
doubling of the $b$ axis \cite{Duc04}. The small intensity of that
scattering might be due to the residual dynamical spin-chain
structure existing down to lowest temperatures. Moreover, the
proposed spin-pentamer state successfully explains the puzzling
controversy between the originally suggested singlet ground state
and the odd number of spins per unit cell in the structure. Note
that in the case observed in doped spin-Peierls and Haldane chains
\cite{Glazkov05,Smirnov02}, these multi-spin objects are diluted and
randomly distributed in the lattice. That gives rise to impurity
induced local magnetic order and phase separation. In contrast,
$\eta$-Na$_{9}$V$_{14}$O$_{35}$ provides a regular lattice of such
spin objects leading the formation of a new exotic ground state.

\section{Summary}
In summary, a model for the magnetic ground state in the quasi-one
dimensional spin system $\eta$-Na$_9$V$_{14}$O$_{35}$ was proposed.
We conclude that the dimerization of spins is far from complete for
$T<10$~K, and the uncompensated vanadium spins provide a constant
contribution to the susceptibility and a linear contribution to the
specific heat. The ground state of $\eta$-Na$_9$V$_{14}$O$_{35}$ can
be understood in terms of "spin pentamers" as building units of a
linear chain along the crystallographic $a$ axis.

\acknowledgments
We thank C.~Helbig, P.~Lunkenheimer and
J.~Hemberger for useful discussions. This work was supported by the
DFG within SFB 484 (Augsburg) and by the Volkswagen-Stiftung.

%%%%%%%%%%%%%%%%%%%%%%%%%%%%%%%%%%%%%%%%%%%%%%%%%%%%%%%%%%

\end{document}